\documentclass[10pt, twocolumn]{article}
\usepackage{graphicx}
\usepackage{subfig}
\usepackage{lipsum}             
\usepackage{blindtext}          

\usepackage[superscript, biblabel]{cite}        

\makeatletter
\renewcommand\@biblabel[1]{#1.}
\makeatother

\usepackage{etoolbox}
\patchcmd{\thebibliography}{\section*{\refname}}{}{}{}

\usepackage[margin=1.0in,hmarginratio=1:1,top=32mm,columnsep=15pt]{geometry}

\usepackage{color}				
\definecolor{dark-gray}{gray}{0.1}	
\usepackage{times}				
\usepackage{microtype}                  
\usepackage[english]{babel}             

\usepackage[font={footnotesize}, labelfont={bf,up}, textfont={sf,up}]{caption}
\usepackage{booktabs}   
\usepackage{mwe}                

\usepackage{enumitem}           
\setlist[itemize]{noitemsep}    

\usepackage{abstract} 
\AtBeginDocument{}                

\usepackage{titlesec}                   
\renewcommand\thesection{\Roman{section}.}                              
\renewcommand\thesubsection{\thesection\Alph{subsection}.}      
\renewcommand\thesubsubsection{\thesubsection\arabic{subsubsection}.} 
\titleformat{\section}[block]{\normalfont\sffamily\bfseries}{\thesection}{1em}{\MakeUppercase}{}        
\titleformat{\subsection}[block]{\normalfont\sffamily\bfseries}{\thesubsection}{1em}{}{}  
\titleformat{\subsubsection}[block]{\normalfont\sffamily\bfseries}{\thesubsubsection}{1em}{}{}  
\titlespacing*{\section}{0.0em}{1em}{0.25em}		
\titlespacing*{\subsection}{0.0em}{1em}{0.25em}	
\usepackage{indentfirst}						

\usepackage{fancyhdr} 

\fancypagestyle{plain}{
\fancyhf{}
\rhead{Submitted to ANS/NT (2021). LA-UR-21-20844}
}

\usepackage{titling}            

\usepackage{hyperref}   
\hypersetup{
        backref=true,
        pagebackref=true,
        hyperindex=true,
        colorlinks=true,
        breaklinks=true,
        urlcolor= black,
        linkcolor=blue,
        bookmarks=true,
        bookmarksopen=false,
        filecolor=black,
        citecolor=blue
}

\pretitle{\begin{center}\large\bfseries}        
\posttitle{\end{center}}                                
\title{\vspace{-0.3in} \sffamily{Canadian Contributions to the
    Manhattan Project and Early Nuclear Research} } 
\author{%
  \normalsize Stephen A. Andrews\thanks{corresponding author: saandrews@lanl.gov},
  Madison T. Andrews and Thomas E. Mason\\[-0.5ex]
  \normalsize Los Alamos National Laboratory \\[-0.5ex]
  \normalsize Los Alamos, NM 87545 } \date{
} 
\renewcommand{\maketitlehookd}{%
\begin{abstract}
\vspace*{-0.5in}
\noindent {\normalfont\textbf{Abstract:}}
\noindent
During the second world war, Canada made several important
contributions to the wartime work of the Manhattan Project. The three
main contributions were: establishing a domestic nuclear research
laboratory in Montreal to investigate heavy water reactors,
creating supply chains to provide uranium oxide, heavy water and
polonium to the Manhattan Project, and the direct contributions of
several Canadians living the United States. These wartime efforts
helped establish a legacy of nuclear research in Canada which has
persisted to the present day.
\end{abstract}
}

\begin{document}

\maketitle

\section{Introduction}

To an outside observer, it may seem strange that Canada has a large
fleet of domestically designed nuclear reactors, joining a small club
of global powers who have achieved such a technological feat. This is
no accident, rather the result of a long history of nuclear research,
reaching back to the start of the 20th century, and a remarkable
partnership between the allied nations to develop a nuclear weapon to
end the second world war. Canadians made three major contributions in
support of this mission: the establishment of nuclear research
facilities in Canada, the delivery of critical raw materials to the
Manhattan Project facilities in the US, and the direct involvement of
several Canadians in the work performed at Manhattan Project
facilities in the US. The following sections will briefly outline
Canada's contribution to these efforts.

\section{Canada's Domestic Research}

\subsection{Pre-war years}

Canada, and in particular the city of Montreal, played an important
part in the very early history of nuclear research when Ernest
Rutherford accepted a position at McGill University from 1899 to
1907. His research at McGill on the transmutations of radioactive
elements earned Rutherford the 1908 Nobel Prize in Chemistry. Otto
Hahn, who would play a pivotal role in the discovery of fission in
1938, joined Rutherford for a time at McGill and investigated
radiation emitted from thorium.

\begin{figure}[t]
  \centering
  \includegraphics[width=\linewidth]{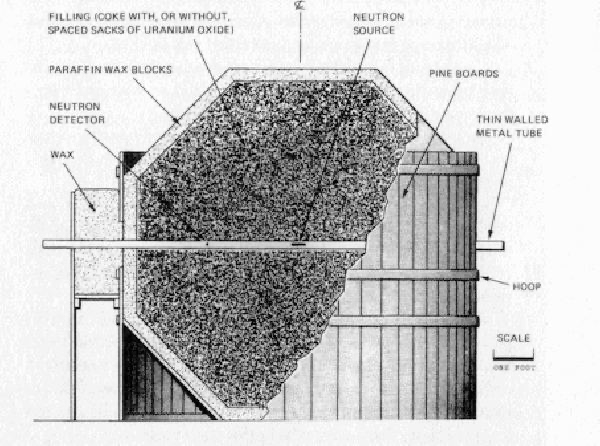}
  \caption{Laurence's multiplication experiment\cite{AECLCanada1997}}
  \label{fig:lawrence}
\end{figure}

Following the discovery of fission and the possibility of a nuclear
chain reaction, Craig Laurence, a scientist at the National Research
Council of Canada (NRC), attempted to construct a nuclear reactor
using natural uranium and a carbon moderator, beginning work in March
1940\cite{LawrenceCanada1947}. The Eldorado mining company was able to
provide a half ton of uranium oxide and he purchased ten tons of
``relatively pure'' calcined coke to act as a
moderator\cite{EgglestonCanada1965}.  Working in the basement of the
Sussex Drive headquarters of the NRC\footnote{just 750 m from 24 Sussex
  Drive, the Prime Minister's official residence}, Laurence arranged the
materials shown in Figure~\ref{fig:lawrence}. The goal of the
experiment was not to achieve a sustained reaction, but rather to
investigate the potential for releasing atomic energy and estimate the
size of a critical reactor\cite{LawrenceCanada1947}. This early effort
predated similar experiments which Enrico Fermi performed at Columbia
University. News of Laurence's work at the NRC led to some degree
of collaboration between the teams in 1941\cite{EgglestonCanada1965}.

\subsection{The heavy water war}

On March 12 1940, two aircraft took off from Oslo, one heading to
Amsterdam, the other to Scotland. The former was forced to land by the
\emph{Luftwaffe}, while the latter continued on with its cargo of a
French officer and almost the entire world's supply of heavy water
($\mathrm{D}_2\mathrm{O}$). Earlier that year, three researchers at
the Coll\`ege de France, Fr\'ed\'eric Joliot, Hans Halban and Lew
Kowarski, had identified this substance as an ideal moderator for
nuclear chain reactions and had submitted several patents related to
the release of nuclear energy\cite{JurgensenFrance2018}\footnote{One
  of these patents ``Perfectionnement aux charges explosives'',
  described the design of a nuclear
  weapon\cite{SabourinMontreal2020}}. The heavy water made its way to
Paris, but before major research could be performed the city was
overrun by advancing German forces. The three scientists and the
precious raw materials were rushed to the port city of
Bordeaux. There, Halban and Kowarski, along with the heavy water and a
few grams of uranium, boarded the \emph{SS Broompark} and set sail for
England on 17 June 1940. Joliot remained behind in France to support
his wife, who was ill\cite{SabourinMontreal2020} and became an active
member of the French resistance\cite{JurgensenFrance2018}. A further 8
tons of uranium oxide was exfiltrated to Morocco by the resistance,
where it remained beyond the reach of Germany for the duration of the
war\cite{JurgensenFrance2018}.

The origins of Canada's wartime nuclear research can be traced back to
this episode. The 185 liters of heavy water would be at the center of
nuclear research in Canada until 1944. While in England, Halban and
Kowarski worked at Cambridge to continue the research which the war
had interrupted. They performed experiments using their natural
uranium samples and heavy water and were able to demonstrate the
feasibility of a divergent chain reaction in
uranium\cite{BretscherReport1940}, however they had far too little
material to begin assembling a working
reactor\cite{SabourinMontreal2020}.

Several months earlier, in early 1940, two other refugee-scientists
Rudolf Peierls and Otto Frisch identified a fast neutron chain
reaction in $^{235}$U as a practical means of developing a nuclear
weapon \cite{RhodesMaking1986}.  This discovery initiated the MAUD
committee in Britain, which investigated the possibility of using
nuclear energy for both weapons and civilian power production. Both
Halban and Kowarski were members of a technical sub-committee, as
they were foreign nationals and could not be full MAUD committee
members\cite{SabourinMontreal2020}. On the 30$^{th}$ of August 1941
Churchill approved the `Tube Alloys' project to begin work on an
atomic bomb. Early on, North America had been identified as an ideal
location for the British project due to its distance from the war, its
proximity to Canadian sources of uranium, and the potential for
collaboration with the American scientific effort. However, due to the
international makeup of the British scientific mission, security
concerns precluded establishing a British research project in the
United States. The British then proposed establishing a nuclear
research facility in Canada, meeting with the president of the
National Research Council of Canada, Dr. C. J. Mackenzie, in August
1942. Mackenzie recognized the importance of this effort to both the
war effort and the potential to get in ``on the ground floor of a
great technological process''\cite{AveryAtomic1995}. Though the
National Research Council's headquarters were located in the national
capital, Ottawa, the new nuclear laboratory was chosen to be located
in Montreal due to its modern airport, large universities, and distance
from the many foreign embassies in Ottawa
\cite{SabourinMontreal2020}. In the autumn of 1942, Halban and the well
traveled heavy water arrived in Canada to establish the Montreal
laboratory.

\subsection{The Montreal Laboratory}

\begin{figure*}[t]
  \centering
  \includegraphics[width=0.75\linewidth]{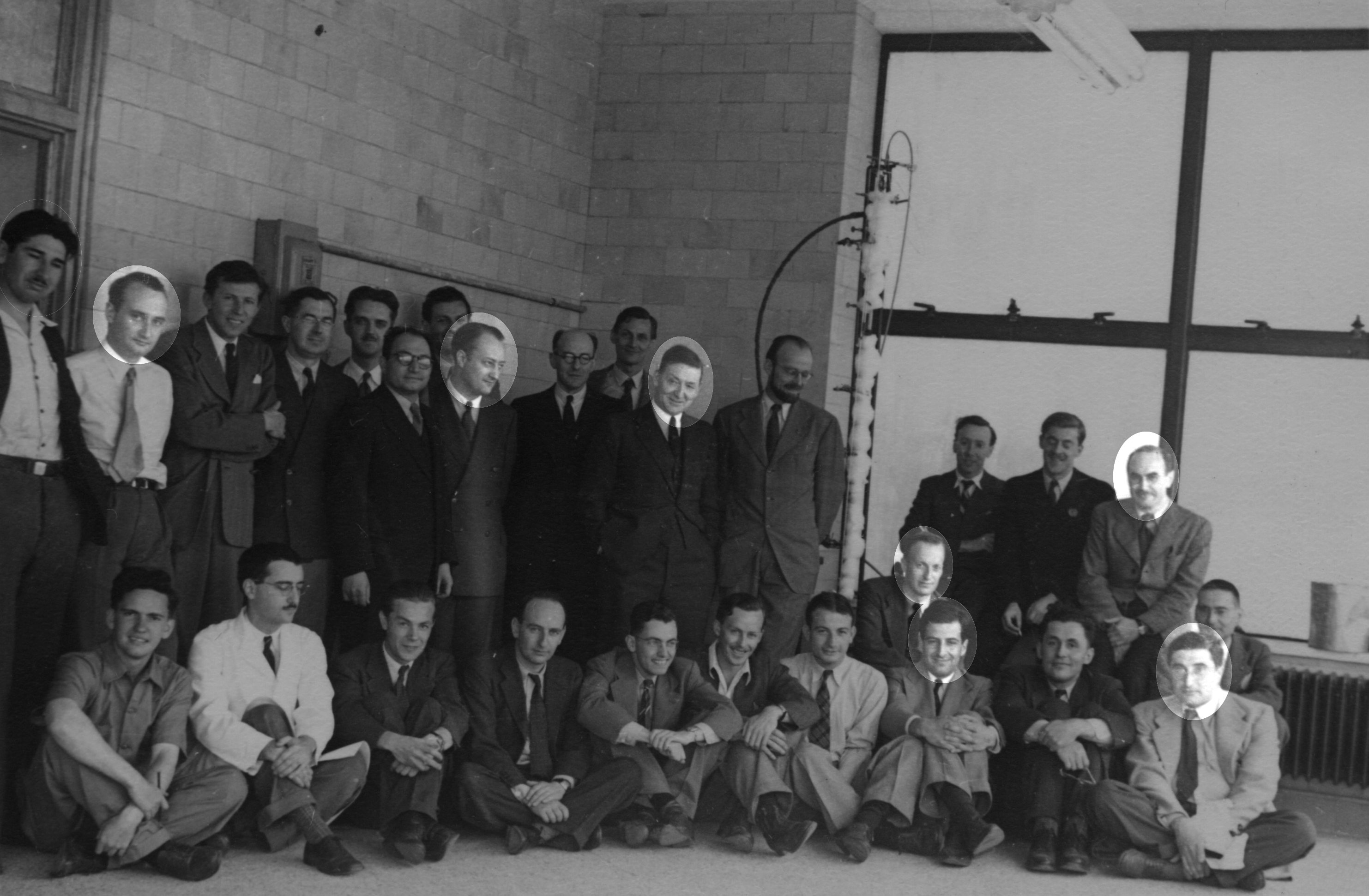}
  \caption{Scientists at the Montreal Laboratory. Highlighted faces
    left to right back: Bertrand Goldschmidt, Hans Halban, John
    Cockcroft, Alan Nunn May, middle: George Laurence, front: Bruno
    Pontecorvo, Georg Placzek. From the archives of the Cockcroft
    family}
  \label{fig:montreal_scientists}
\end{figure*}

\begin{figure*}[t]
  \centering
  \subfloat[View of the Ontario-Quebec region]{
    \includegraphics[width=0.49\textwidth]{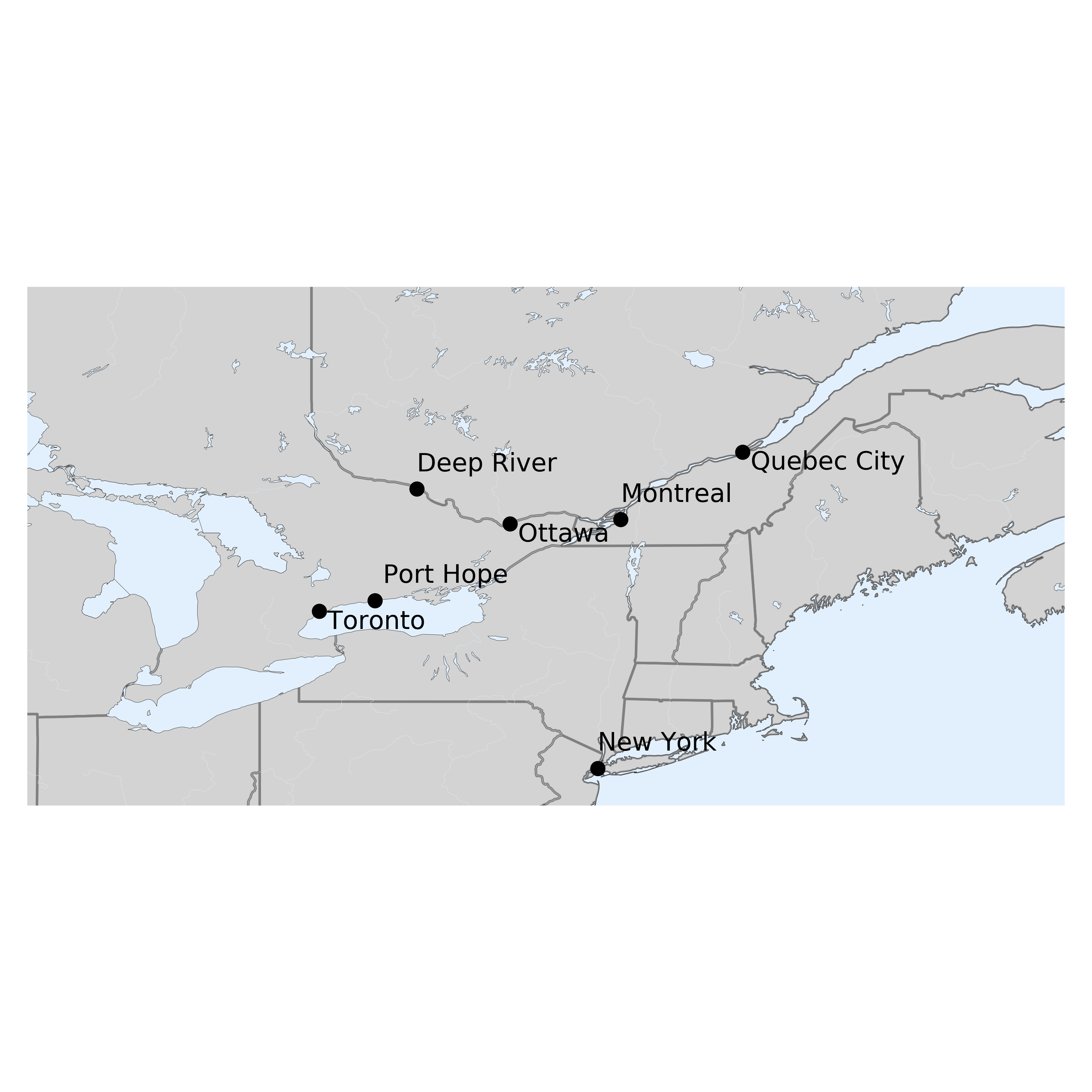}}
  \subfloat[View of Canada]{
    \includegraphics[width=0.49\textwidth]{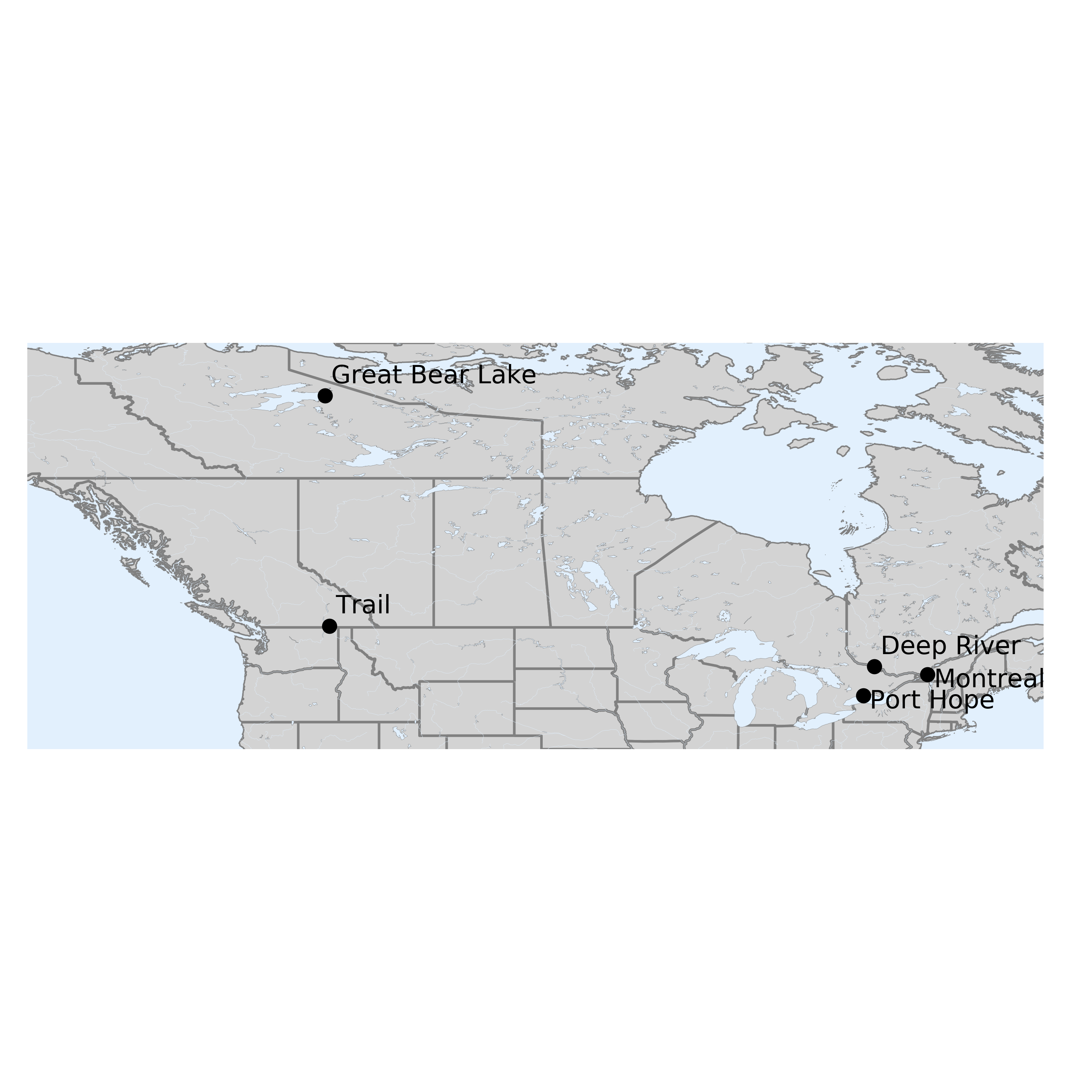}
  }
  \caption{Maps of the important areas of Canada for wartime nuclear
    research and production.\label{fig:maps}}
\end{figure*}

The first set of researchers, including Halban, arrived in Montreal on
the 24$^{th}$ of September 1942. As the leading authority on heavy
water moderated nuclear reactions, Halban was chosen as the director
for the research group. They initially rented a large house on Simpson
Street, and all available space was quickly converted into offices for
the researchers. The University of Montreal offered a wing of their
newly completed medical building, and this would become the home of
the Montreal Laboratory for the duration of the project. Absent from
the initial team was Kowarski, who remained in Cambridge. Though he
had worked productively with Halban for many years as a colleague,
Kowarski rejected the possibility of being his subordinate in
Montreal\cite{SabourinMontreal2020}. A photograph of the main
researchers at the Montreal Laboratory is shown in
Figure~\ref{fig:montreal_scientists}, and maps showing important
locations for Canadian wartime nuclear research are given in
Figure~\ref{fig:maps}.

Soon after the laboratory was set up, however, the project almost
collapsed. In December 1942, the United States ceased almost all
cooperation with the British on matters other than isotope
separation. This cooling of relations happened after Fermi's graphite
moderated reactor worked, so the heavy water research efforts of
Montreal were less critical to the success of the Manhattan
project. Without access to heavy water or uranium, the work of the
Montreal Laboratory was limited to small-scale investigations using
only the heavy water evacuated from France.

Despite the temporary freeze in communication, many useful research
efforts were undertaken in 1943-1944. Many technical reports were
published and shared with the Manhattan Project, mostly on topics
related to neutronics \cite{PlaczekFunctions, PlazekAngular1943,
  WallaceElementary1943, PlaczekNeutron1943, MarkMilne1943,
  MarshakMilne1944, MarkNeutron1944, MarkSpherical1944,
  SeidelInfluence1946, ClaytonCritical1946, ClaytonFate1946}. These
reports are still available at Los Alamos, from the National Security
Research Center archives. Additionally, the work that Halban and
Kowarski had undertaken in Cambridge was continued to further
investigate the properties of a homogeneous reactor using a slurry of
uranium and heavy water. Experiments on a carbon-moderated pile were
also performed\cite{EgglestonCanada1965}.

During this time, a French scientist working in Montreal, Bertrand
Goldschmidt, who had previously worked with Seaborg and others on
problems of plutonium separation at the Metallurgical Laboratory in
Chicago, returned to visit his friends at that lab. Having never
handed in his badge, his friends let him ``borrow'' four micrograms of
plutonium\cite{SabourinMontreal2020} and a small quantity of fission
products from which he was able to extract a further three micrograms
of plutonium\cite{GowingBritan1964}. These allowed Goldschmidt to
begin studying plutonium and methods of separating plutonium from
other fission products while working in Montreal.

Despite the progress made at the Montreal Laboratory, the lack of
cooperation between the Canadian/British and American efforts had to
be addressed in order for the Canadian effort to continue. In August
1943, the three national leaders, Winston Churchill, Franklin
D. Roosevelt and Canada's Prime Minister William Lyon Mackenzie King,
met at the citadel in Quebec City to coordinate the war effort. Among
the many pressing topics to be resolved at this conference, an agreement on
nuclear research was reached to allow full cooperation in the field of
scientific research, and sufficient communication in more applied
fields to, ``quickly bring the project to
fruition.''\cite{EgglestonCanada1965} Additionally, the three
governments agreed never to use nuclear weapons against each other,
never to use them against third parties without the others consent and
not to communicate information on nuclear research to third parties
without mutual consent\cite{EgglestonCanada1965}.

Despite the agreement between the national leaders, full cooperation
between the Montreal Laboratory and the Manhattan Project could not
proceed without the support of Gen. Groves. In March 1944, Groves
and James Chadwick, one of the leaders of British atomic research, had
a critical meeting. Ultimately, they agreed to proceed with the
construction of a Canadian heavy water moderated reactor in light of
the technical advantages of the heavy-water design and potential for
post-war uses. It was acknowledged that the design and construction of
this plant would not be completed in time to make an impact on the
immediate plutonium production needs of the Manhattan
project\cite{GowingBritan1964}. This agreement also made sufficient
quantities of heavy water and uranium available for the reactor
\cite{ManhattanBookIVol4Ch9}.

An additional requirement of the improved cooperation between the
Montreal Laboratory and the Manhattan Project was a change of
leadership. Though Hans Halban had lead the laboratory since its
inception, he was not a natural leader, and Gen. Groves insisted that a
British scientist lead the Montreal
laboratory\cite{SabourinMontreal2020}. So, in April 1944, John
Cockcroft, a British scientist who had already made important wartime
contributions to the radar project, arrived in Montreal to assume the
leadership of the laboratory and begin a new, more active phase of
research.

\subsection{Move to Chalk River}

With a clear plan to build a heavy water reactor, the Montreal
laboratory needed a larger site to build the plant and the required
laboratories. The requirements for the site were similar to those of
Los Alamos, in particular, a remote location was desired, but with
good access to transportation in order to bring in
supplies. Additionally, abundant supplies of freshwater were needed to
cool the reactor. A site on the Ottawa River, almost 200 km upstream
from the national capital was chosen. The new town of Deep River was
established nearby in order to provide housing for the staff and
scientists. Unlike the somewhat organic growth
of Los Alamos, a professor of urban planning at the University of
Montreal led the design of the new town \cite{EgglestonCanada1965}.

\begin{figure*}[t]
    \centering \subfloat[Zero Energy Experimental Pile (ZEEP): The
    first reactor to go critical outside the U. S. \label{fig:zeep}]{
      \includegraphics[width=0.45\textwidth]{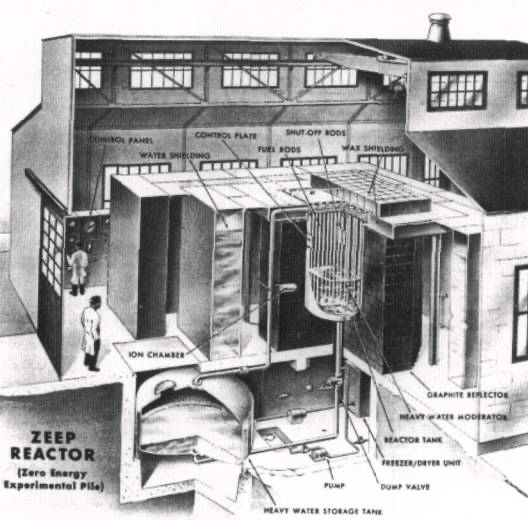}}
    \subfloat[Chalk River in 1954. The low white building in the
    foreground contained ZEEP.  The adjacent large brick building
    contained the NRX reactor, The NRU reactor, a post-war project, is
    under construction \label{fig:nrx}]{
      \includegraphics[width=0.45\textwidth]{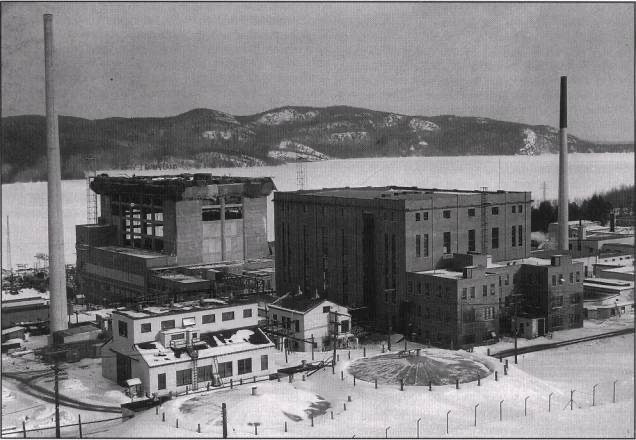}}
    \caption{The early reactors at Chalk River}
    \label{fig:cnl_reactors}
  \end{figure*}

The site for the town and laboratory was chosen in mid July 1944, and
construction of the laboratory was begun by the crown corporation
Defense Industries Limited\cite{EgglestonCanada1965}. The main focus
of the laboratory was to begin construction of the heavy water
reactor, termed the National Research eXperimental reactor (NRX). It
was to have a 10 MW power output\cite{ManhattanBookIVol4Ch9}, and
produce 7 g / day Pu and 0.5 g / day $^{233}$U, as well as other
isotopes for research and medical applications. The design of the
reactor was complex, with aluminum sheathed fuel, cooled by purified
river water flowing through annular channels around the clad
fuel. These assemblies were immersed in the heavy water moderator. The
shielding surrounding the calandria contained channels where rods of
thorium could be exposed to the intense neutron flux to allow the
production of $^{233}$U. A view of the reactor building is shown in
Figure~\ref{fig:nrx}.

This design presented significant technical challenges, and it was
proposed that a simple Zero Energy Experimental Pile (ZEEP) be
constructed to test optimal arrangements of fuel and moderator. A
drawing of this reactor is shown in Figure~\ref{fig:zeep}. This
reactor diverted 5 tons of $\mathrm{D}_2\mathrm{O}$ and 3.5 tons of
uranium metal from the main reactor\cite{ManhattanBookIVol4Ch9}, but
due to the lengthy design and construction timelines for the NRX
reactor, this did not slow down the overall program. Lew Kowarski
arrived from Cambridge, ready to lead work on this new reactor now
that Halban was no longer the lab director. On Sept 5 1945, ZEEP was
the first reactor outside the United States to become critical
\cite{EgglestonCanada1965}.

ZEEP went critical two days after Japan's official surrender and the
end of the Second World War. The UK then focused on its own domestic
nuclear program at Harwell, and John Cockcroft was recalled to England
to lead these efforts. He was succeed by W. B. Lewis as the director
of the laboratory. Lewis would continue to serve in various senior
leadership roles at the laboratory until 1973\cite{FawcettNuclear1994}
overseeing the completion of the wartime projects and leading Canadian
research efforts in the nuclear era.











\subsection{Post-war efforts}

Due to the complexity of the reactor design and the lack of wartime
pressures on its development, it took a further year after the end of
the war for the main reactor, named NRX, to go critical, which it did
on July 22 1946.


A parallel research effort to the development of the reactors was
developing facilities to process the irradiated fuel to obtain
$^{239}$Pu and $^{233}$U from the irradiated uranium and thorium,
respectively\cite{EgglestonCanada1965}. Despite the cooperation
between the Montreal Laboratory and the Manhattan Project on reactor
design and materials, research relating to chemical separation was
strictly controlled by the Americans. Scientists at the Montreal
laboratory had to develop their own processes to separate the
desired isotopes from the highly radioactive fuel. They developed a
process which used triethylene glycol dichloride to separate out the
$^{239}$Pu, and a different process was used for $^{233}U$
separation\cite{ManhattanBookIVol4Ch9}.  The production of $^{239}$Pu
began in August 1949 and of $^{233}U$ on January 1951.


\subsection{Security}

The Canadian efforts in support of the Manhattan Project in Montreal
and Chalk River were targets for Soviet espionage, as were the British
and Americans efforts. Bruno Pontecorvo and Alan Nunn May were the
primary sources, but  the technical information they revealed was
somewhat removed from the design efforts at Los Alamos, and therefore
of lower value.  A much more significant impact came with the
defection of Igor Gouzenko from the Soviet embassy in Ottawa on
September 5, 1945 – three days after the end of the war.  This became
known as the ``Gouzenko Affair.''  The facts surrounding the defection
became known following the defection through a Royal Commission
established by Prime Minister Mackenzie King\cite{CloutierRoyal1946},
Gouzenko’s own autobiographical account\cite{GouzenkoIron1948}, and
a film of the same name.  A more complete account became possible in
2003 when the Canadian Security Intelligence Service and MI5
declassified the files, and the collapse of the Soviet Union provided
insight into the ramifications for the Soviets\cite{KnightHow2005}.

Igor Gouzenko was a cipher clerk for the GRU (Soviet foreign military
intelligence) working at the Soviet embassy in Ottawa from June 1943
until his defection in September 1945.  He was responsible for
encrypting and decrypting communications between Ottawa and Moscow,
giving him access to all aspects of Soviet intelligence gathering.
Gouzenko and his wife made the decision to defect about a year before
the end of the war when he received news that he was to be transferred
home to Moscow. The loss of a much more comfortable and free life they
had grown accustomed to, as well as the possibility of more severe
punishments for those called home, led him to begin collecting
documentation that he could use to confirm his value as a defector.
His recall was postponed for a year due to the difficulty of replacing
him, giving him time to collect 250 pages of documentation to support
his disclosures. He was aware that Victor Kravchenko, a previous
defector without hard evidence, had difficulty convincing the FBI of
his veracity.  Initially, he was rebuffed by the Ottawa Journal
newspaper and the Department of Justice. After a bizarre sequence of
events with NKVD agents ransacking the Gouzenkos apartment, the Ottawa
police assisted them in making contact with the Royal
Canadian Mounted Police.  The initial reaction from the Canadian
authorities, including the Prime Minister, was not welcoming, as they
feared the potential damage to relations with the Soviet Union, which
was still formally a wartime ally.  Eventually, the Canadians decided
to debrief Gouzenko, and by September 12 had contacted the FBI leading
Director, J. Edgar Hoover, to send a letter to the White House with
the revelation that ``the obtaining of complete information regarding
the atomic bomb the Number One project of Soviet
espionage.''\cite{HooverHoover1945} The supporting documentation
described the theft of a sample of $^{233}\mathrm{U}$ by Alan Nunn
May, which had been hand delivered to Lavrentiy Beria\footnote{head of
  the NKVD} and a Soviet spy at high levels in the U.S. State
Department who became mistakenly identified as Alger Hiss.

The impact of these revelations in Ottawa, Washington, and London were
significant.  It led to arrests, convictions, and a concerted effort
to uncover the full extent of Soviet espionage that revealed Klaus
Fuchs, the Rosenbergs, and ultimately the McCarthy hearings.  It
marked the beginning of the transformation of the Soviet Union from
wartime ally to geopolitical adversary.  The implications of a breach
of this magnitude in the Soviet Union were also substantial – the
Soviets were very quickly and completely informed of what Gouzenko had
revealed, through a highly placed Soviet spy in British
counterintelligence, Kim Philby.  Ironically, the Soviet espionage
efforts targeting Canada had not been successful in revealing
information of direct relevance to weapons design.  The most
significant asset was Nunn May, who had been recruited while he was in
the UK.  John Cockcroft, who was leading Chalk River at the time, was
shocked when he learned on September 10, 1945 that Nunn May was a
spy. In 1947 he was among the scientists signing a petition advocating
a reduction in the 10 year sentence Nunn May received, although he
later expressed regret for that act\cite{HartcupCockcroft1984}. Bruno
Pontecorvo has long been suspected of being a Soviet spy, possibly
revealing details of the design of the NRX reactor\cite{CloseHalf2015}.
He maintained he had never worked on weapons nor transmitted secrets
but his defection to Russia following his return to the UK suggests
otherwise. Efforts to recruit Canadian sources were generally not so
successful, as the scientists targeted for their communist sympathies
were either distant from the Manhattan Project activities or unwilling
to transfer information that was not already available to the Soviets.

\section{Raw materials}

\subsection{Uranium mining and milling}

The town of Port Hope sits on the shores of lake Ontario, 100 km east
of Toronto. Today, this small town is home to a large uranium fuel
production facility for Canada's CANDU reactor fleet. In the late
1930's, however, this was one of the few facilities in the world that
could process pitchblende ore into uranium oxide, and it would play an
important role in supplying the Manhattan Project with raw
materials. In 1932, a rich deposit of pitchblende was discovered on
the shores of Great Bear Lake in the Northwest Territories, and was
mined by Eldorado Gold Mines, Ltd. The commercial interest in this
mine was not the uranium, but rather the radium contained in the
ore. The refinery in Port Hope was built to extract the radium,
beginning operations in 1933. Due to the outbreak of the war and the
challenges of obtaining supplies to operate a mine in the arctic,
mining operations ceased in July 1940. In 1941, the company was
approached by a member of the American advisory committee on uranium,
Dr. Lyman Briggs, who wished to purchase uranium oxide, which had been
a byproduct of the radium extraction. Subsequent purchase orders for
uranium oxide from the United States led to the Great Bear Lake mine
reopening, resuming operations in August
1942\cite{EgglestonCanada1965}. As the strategic importance of uranium
was becoming clearer to those in the industry, Canadian Minster of
Munitions C. D. Howe purchased a majority share in Eldorado in 1943
for the British and Canadian governments\cite{GowingBritan1964}, to
keep prices under control.

\begin{figure}[t]
  \centering
  \includegraphics[width=\linewidth]{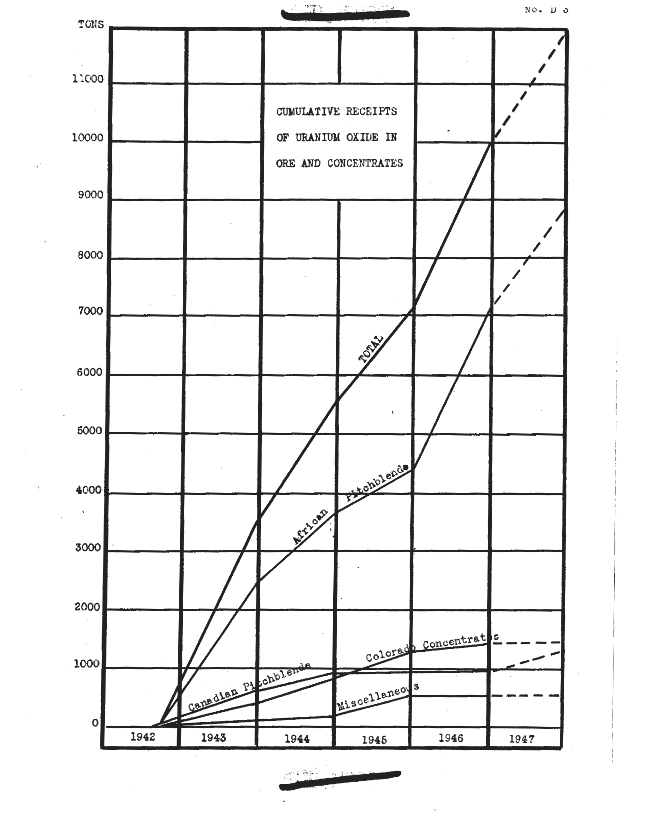}
  \caption{\label{fig:ore_breakdown} Contribution of uranium bearing
    ores from various sources. Note that a large fraction of the
    African ores were processed in Port Hope\cite{ManhattanBookVII}}
\end{figure}

Canada was not the only source of uranium to the Manhattan
project. The Union Mini\`ere du Haut Katanga had been producing radium
from its particularity rich deposits of pitchblende in the Shinkolobwe
mine in the Belgian Congo. They had supplied the team at the Coll\`ege
de France with the uranium they subsequently hid from the Germans. In
1940, as war in Belgium seemed likely, the director of the company,
M. Edgar Sengier, had all available ore in the Belgian Congo shipped
to New York. This ore was sold to the US government. However, as
shipments from Great Bear Lake had yet to pick up, the processing
facilities in Port Hope were available to process the pitchblende ore
into uranium oxide. Port Hope refined all 1200 tons of the Congolese
ores\cite{GrovesNow1963, GowingBritan1964}. Subsequent shipments of
ore arrived from the Belgian Congo during the war and were refined
both at Port Hope and by the Vitro Manufacturing Company in
Pennsylvania\cite{ManhattanBookVII}. A list of the sources of uranium
ores used by the Manhattan Project is shown in
Figure~\ref{fig:ore_breakdown}.
The uranium oxide
from all sources was combined together for subsequent processing, so it
is impossible to trace the exact fate of the ores processed in Canada,
but for the first three years of the Manhattan Project, ores processed
at Port Hope made up the majority of the uranium oxide used in the
Manhattan Project. It is very likely that some significant part of the
uranium refined to pure $^{235}$U for the Little Boy weapon, as well
as the uranium which fueled the research reactors and Hanford reactors,
passed through Port Hope. A more detailed analysis of the flow of
materials through the Manhattan Project suggests that the majority of
the uranium extracted from Great Bear Lake would have been used in
research reactors rather than as inputs to the isotope separation
plants or Hanford reactors\cite{ArsenaultManhattan2008}.

\subsection{Heavy Water}

The uranium mining and milling operations were not Canada's only
contribution of raw materials to the Manhattan Project. The
Consolidated Mining and Smelting ammonia plant in Trail, British
Columbia \cite{ManhattanBookIII} was modified to produce heavy water,
in addition to its normal output of fertilizer. Trail is situated on
the Columbia River in a region with huge potential for hydroelectric
power, a location geographically similar to the heavy water facility
in Norway. The Corra Linn Dam on the nearby Kootenay River provided
the power required for ammonia production\cite{CanadaBalance2019}. The
American government negotiated directly with Consolidated Mining and
Smelting to modify the plant and purchased its entire
output~\cite{EgglestonCanada1965}. The first deliveries of heavy water
began on 16 June 1943 at a rate of 35 lb / month and full production
of 1000 lbs / month was achieved December 1944\cite{ManhattanBookIII}. The
US also established heavy water plants at Morgantown, WV Ordnance
Works; Wabash River, IA Ordnance Works; and Alabama Ordnance
works\cite{ManhattanBookIII}, but the plant at Trail was the first to
begin operations.


\subsection{Polonium}

A final strategic material for the Manhattan Project, which Canada
was able to supply, was polonium. This was used to create neutron
sources used in the initiators for both the uranium and plutonium
weapons. Two Montreal Laboratory scientists Fritz Paneth and
Bertrand Goldschmidt met with  Manhattan Project scientists to discuss
methods of extracting polonium from lead
dioxide\cite{HoddesonCritical1993}. Several tons of lead dioxide were
in storage at the Port Hope refinery, a byproduct of the uranium
milling activities. This ``radioactive lead'' was purchased from
Eldorado \cite{ManhattanBookVII} and shipped to Monsanto facilities in
Dayton, Ohio, where it supplemented polonium produced by irradiated
bismuth. The majority of the polonium used by the Manhattan Project
was eventually produced from the irradiated bismuth, but  Port Hope
provided an important source of polonium in the early days when it was
less clear that the bismuth process would work~\cite{HoddesonCritical1993}.

\section{Canadians In the Manhattan Project}

\begin{figure*}[t]
  \centering
  \subfloat[Robert Christy]{\includegraphics[width=0.35\textwidth]{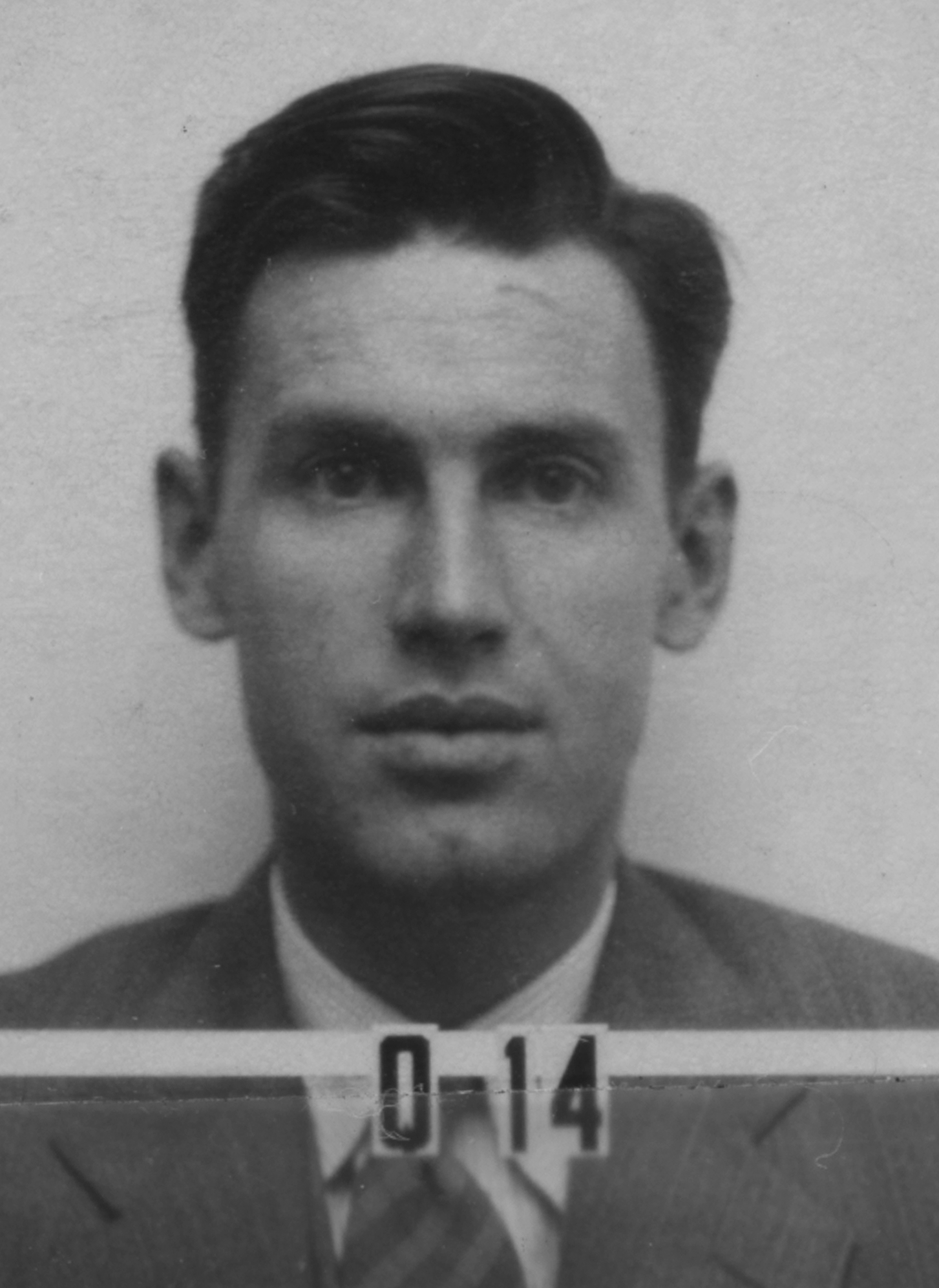}}
  \subfloat[Louis Slotin]{\includegraphics[width=0.35\textwidth]{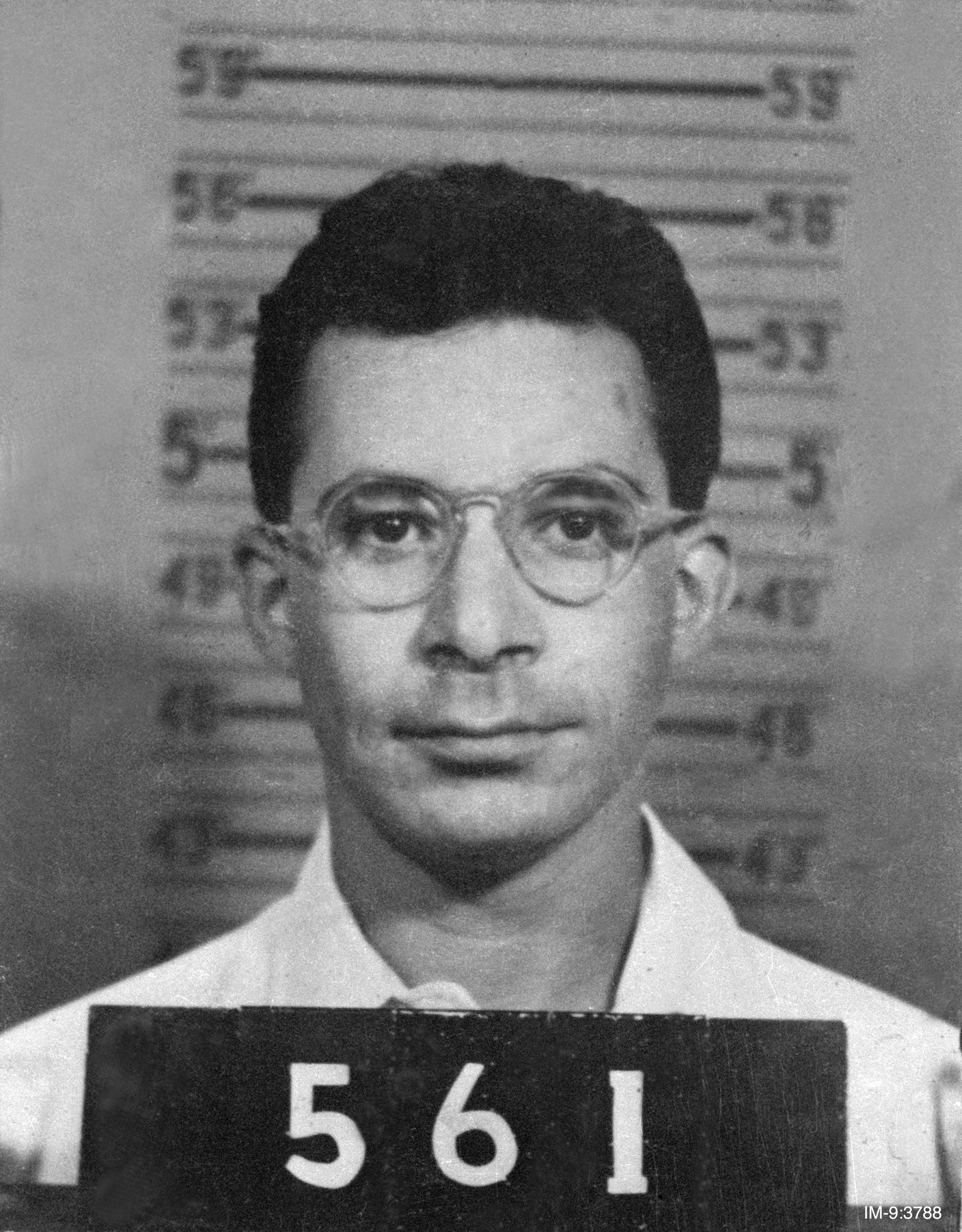}}\\
  \subfloat[Walter Zinn]{\includegraphics[width=0.35\textwidth]{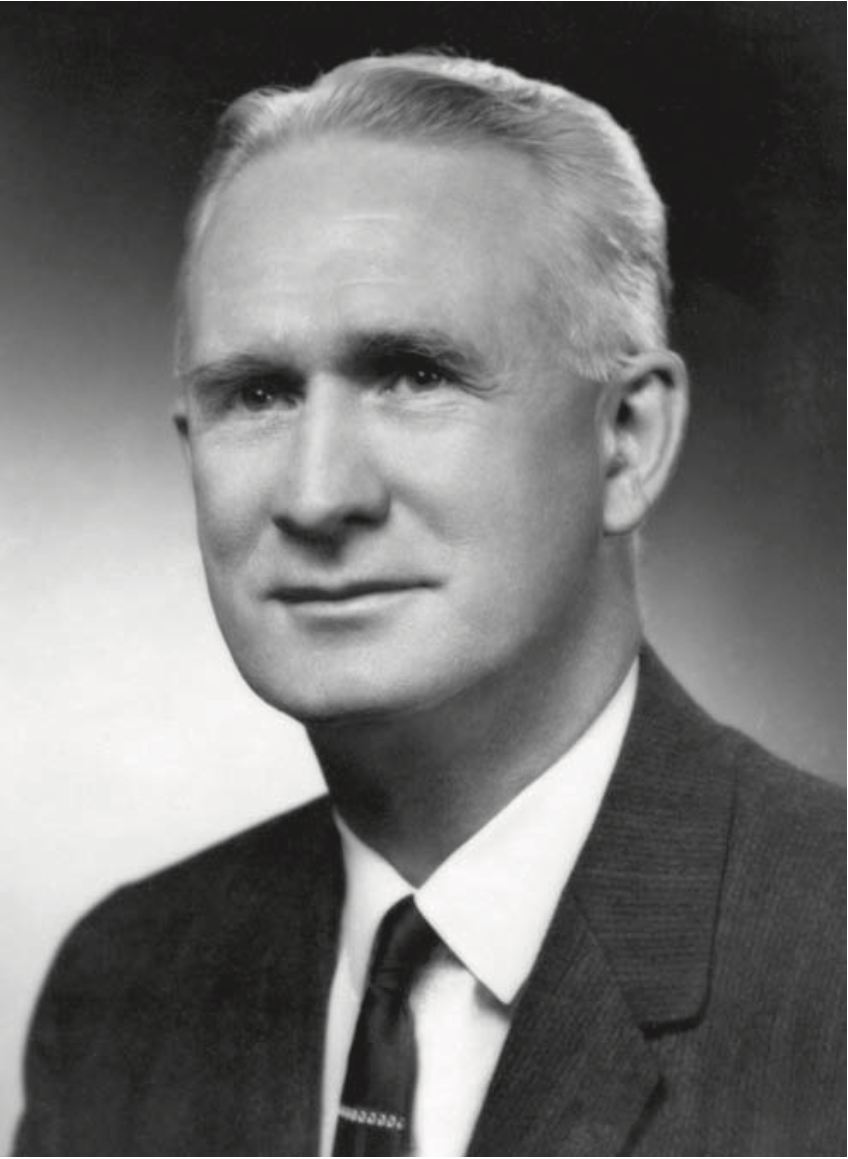}}
  \subfloat[Carson Mark]{\includegraphics[width=0.35\textwidth]{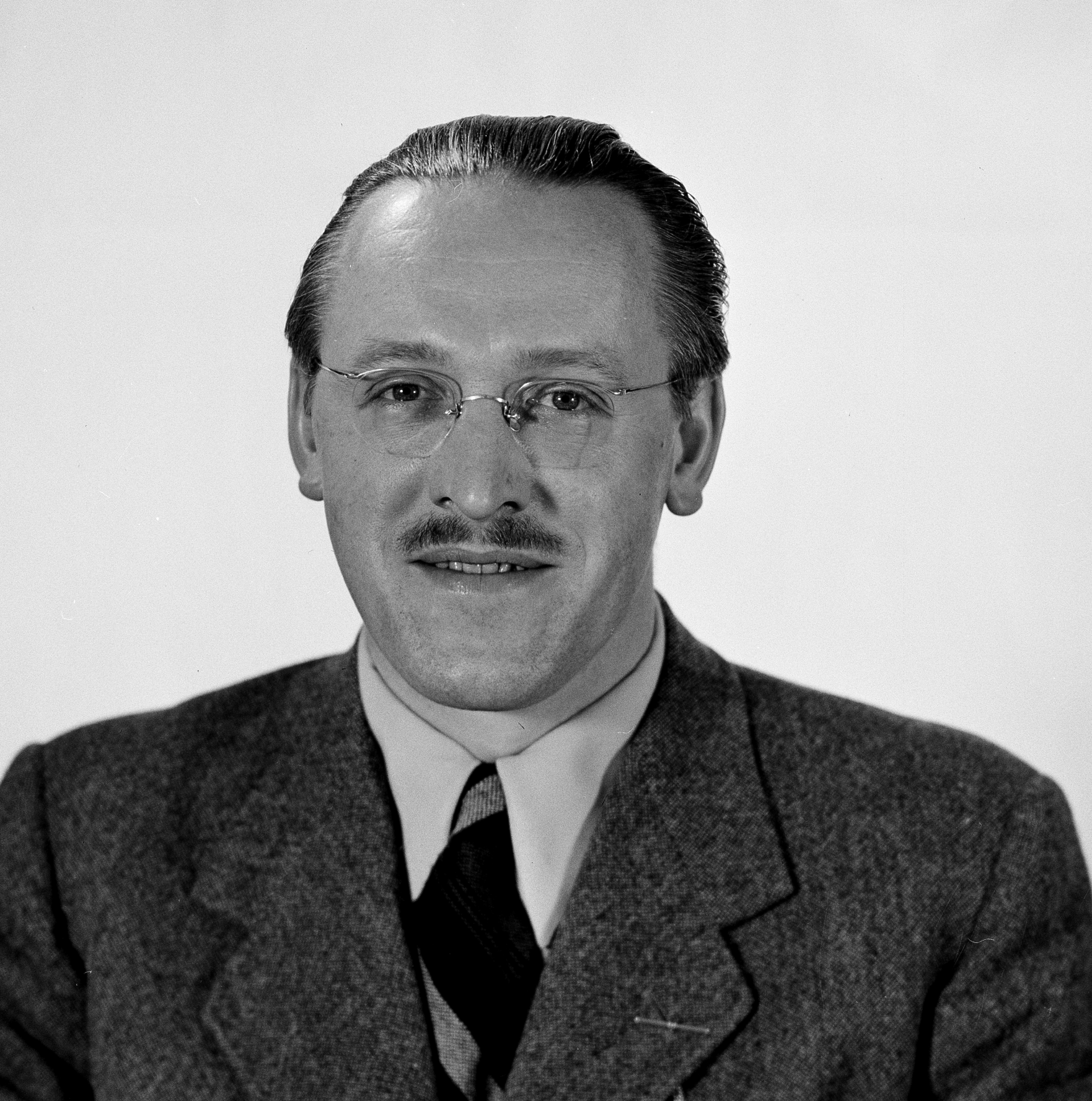}}
  \caption{Notable Canadians working in the Manhattan Project  \label{fig:canadians}}

\end{figure*}

In addition to Canada's domestic wartime research and raw materials
production, several Canadian-born scientists worked directly for the
Manhattan Project. Generally, these scientists were graduate students
or professors at American universities at the outbreak of the war. In
the pre-war years, opportunities for graduate degrees in Canada were
limited, so it was common for students to pursue their studies in the
US or UK. Table~\ref{tab:canadians} shows a list of all the wartime
members of Los Alamos who had Canadian heritage. A selection of
Canadian scientists who had a particularly significant contribution to
the Manhattan Project are discussed in greater detail in the following
sections, and their portraits are shown in Figure~\ref{fig:canadians}.

In addition to the Canadians who worked on the Manhattan project,
several foreign scientists transferred from the Montreal Laboratory to
Los Alamos near the end of the war. Notable among them were Georg
Placzek and Bengt Carlson.  Placzek was a Czech scientist who worked
with Bohr in Copenhagen\cite{RhodesMaking1986}, moving to Columbia
University where he collaborated with Fermi on reactor
design\cite{RhodesMaking1986}. He then arrived in Montreal to lead the
Theoretical Physics Division\cite{EgglestonCanada1965}, where he
published prolifically on neutronics \cite{PlaczekFunctions,
  PlaczekNeutron1943, PlazekAngular1943}. Placzek arrived in Los
Alamos in mid 1945 and was made group leader of T-8 ``Composite
Weapons''. Bengt Carlson was a Swedish physicist who worked for the
Montreal Laboratory in 1943. Like Placzek, he arrived in Los Alamos in
1945 and worked in the Czech scientist's group. Carlson remained at
Los Alamos until his retirement in 1976\cite{ABQBrengt2007}. Carlson
was heavily involved in the development of computational physics at
Los Alamos and made significant contributions to the field of
neutronics by developing the discrete ordinate, `Sn', method for
modeling neutron transport.

\begin{table}[t]
  \caption{Canadians working at Los Alamos during the Manhattan Project}
  \label{tab:canadians}
  \begin{tabular}{ll}
    \hline
    \hline
    James S. Allen & G-Division\\
    Hugh G. Bryce & X-Division\\
    J. Carson Mark & T-Division\\
    Robert Christy & T-Division\\
    John Gaston Fox & X-Division\\
    Roy W. Goranson & G-Division\\
    Hyman Rudoff & CM-Division\\
    Charles Edward Runyan & X-Division\\
    Manuel Schwartz & T-Division\\
    Louis Slotin & G-Division\\
    Roger B. Sutton & R-Division\\
    John H. Williams & P-Division\\
    Lester Winsberg & F-Division\\
    \hline
    \hline
  \end{tabular}
\end{table}

\subsection{Robert Christy}

Robert Christy was a scientist at the Metallurgical laboratory in
Chicago and at Los Alamos. He made important contributions to the
design of the implosion weapon. Christy was born in Vancouver, British
Columbia in 1916 and studied at the University of British Columbia
before beginning a Ph.D. program at Berkeley under the mentorship of
J. Robert Oppenheimer.

After graduate school, he joined the faculty of Illinois Tech and then
joined the Metallurgical laboratory in the University of Chicago in
1942. Christy assisted in machining the graphite for the pile and was
present during the startup of the CP-1
reactor\cite{ChristyInterview1994}.

In early 1943, Oppenheimer recruited Christy to come and work in Los
Alamos, arriving around March, accompanied by his wife. The couple
initially shared an apartment in Pojoaque with Richard
Feynman\cite{FeynmanSurely1985}. Christy worked in the Theoretical (T)
division, and his experience making calculations on fast neutron
systems was first used to estimate the critical mass of the
homogeneous $^{235}U$ reactor, named the ``water
boiler.''~\cite{MeyerWater2021LANL} When the reactor was assembled,
Christy's `pre-shot' predictions agreed very closely with the measured
values, enhancing the confidence in T Division's
calculations\cite{HoddesonCritical1993}. Christy also made important
contributions to the design of the trinity ``gadget'' and this is the
subject of another paper in this special
issue\cite{ChadwickInvented2020}.




\subsection{Louis Slotin}

Louis Slotin was a scientist who worked at Los Alamos during the war,
but is more well known for his involvement in a criticality accident in
the post-war years.

Slotin was born in 1910 in Winnipeg, Manitoba and educated at the
University of Manitoba. He then traveled to England to attend King's
College London, where he earned both a Ph.D. and the title of King's
College amateur bantam-weight boxing championship in 1936. He began
work at the University of Chicago in 1937, working on a cyclotron. He
joined the Metallurgical Laboratory when it was established at the
University of Chicago, and, according to university records, he was
also present during the first run of the CP-1 reactor. Slotin arrived
in Los Alamos in Dec 1944, having previously worked at Oak
Ridge\cite{ZeiligLouis1995}. Slotin's high tolerance for personal risk
was demonstrated during his time at Oak Ridge. On a Friday afternoon,
Slotin wanted to make repairs to an experiment, which was at the
bottom of a tank of water used as a biological shield for a
reactor. Over the weekend, Slotin independently decided to swim in the
reactor pool and make the repairs underwater
\cite{ZeiligLouis1995}. Experiments were able to proceed on Monday,
but Slotin's lax\cite{ZeiligLouis1995}
attitude to experimental safety would lead to tragedy as he continued
his Manhattan Project work at Los Alamos.

At Los Alamos, he participated in assembling plutonium weapons and
lead the assembly efforts for the core of the trinity device. He
became the lead for all Plutonium experiments and was group leader for
M-2, `Critical Assemblies'~\cite{FromanProgress1945}. His expertise in
working with such assemblies led to the criticality accident which
claimed his life on 21 May 1946. Slotin was performing an experiment
where he used a screwdriver to lower a beryllium hemisphere onto a
6.19 kg sphere of nickel plated plutonium\cite{HankinsRevised1968} in
order to demonstrate concepts of criticality, including neutron
reflection. As the beryllium was lowered, Slotin's grip on the
screwdriver slipped and the two hemispherical beryllium reflectors
closed. Though Slotin acted quickly, separating the assembly and using
his body to shield others in the room, he had already received a
lethal dose of radiation\cite{ZeiligLouis1995}. Other observers also
received significant doses of radiation, though they all
recovered. Slotin died nine days later, on 30 May. Gen. Groves
arranged a special military flight to bring Slotin's parents to his
bedside before he died. Before his death, Gen. Groves also sent Slotin
a letter commending him for his ``bravery and quick action which saved
the lives of seven co-workers.''\cite{ZeiligLouis1995}


\subsection{Walter Zinn}

Walter Zinn was not a scientist at Los Alamos, but was a key figure in
the Metallurgical Laboratory in Chicago, and later became the first
director of Argonne National Laboratory. Born in 1906 in
Berlin,~\footnote{Renamed Kitchener during WWI, for patriotic reasons
  during the war} Ontario, Zinn attended Queen's university for his
undergraduate degree, and received a Ph.D. from Columbia in
1934\cite{WeinbergZinn2004}. He worked with Leo Szilard and Enrico
Fermi at Columbia on their early work on fission, and moved to Chicago
to work with Fermi on the development of CP-1, the world's first
nuclear reactor. Zinn was present at the start up of CP-1 and he had
the most important job of the several Canadians present for this
historic event. He was the `axe man' responsible for cutting the rope
which kept a spring-loaded rod of cadmium out of the reactor in the
event of an uncontrolled reaction\cite{WeinbergZinn2004}. In 1946,
when the Manhattan Project was transferred to the Atomic Energy
Commission, Zinn remained in the US as the first director of
Argonne\cite{WeinbergZinn2004}, though was considered, alongside
W. B. Lewis, as a replacement for Cockcroft as the director of the
Montreal Laboratory \cite{EgglestonCanada1965}.

\subsection{J. Carson Mark}

Carson Mark was a Canadian scientist who worked at the Montreal
Laboratory for the majority of the war, but moved to Los Alamos with
other commonwealth scientists in the spring 1945. Carson Mark was born
in 1913 in Lindsay, Ontario, educated at the University of Western
Ontario, and received a Ph.D. from the University of Toronto in
1935\cite{PetschekCarson1997}. In 1947, he became division leader for
T-Division, a position he kept until his retirement in
1973\cite{LANLCarson2012}. During his time at the Montreal Laboratory, he
published several technical reports, including work on neutron flux
near plane surfaces,~\cite{MarkNeutron1944} as well as applying the
spherical harmonic methods to transport
problems~\cite{MarkSpherical1944}. At Los Alamos, Carson led
T Division through the period where thermonuclear weapons were
developed. As division leader, he was effective at the
challenge of managing the strong
personalities in the division. He also was able to:

\begin{quotation}
  encourage free scientific pursuits in areas which were only
  indirectly related to the tasks of the laboratory, and he[Carson Mark] supported
  theoretical physics and applied mathematics in the best
  sense\cite{UlamAdventures1976},
\end{quotation}

creating a tradition of scientific excellence that has continued to
the present day.

\section{Conclusion}

During the second world war Canada made three significant
contributions to the work of the Manhattan Project: a domestic
nuclear research program, the production of essential raw materials,
and the contributions of Canadian scientists working directly for the
Manhattan project. In addition to helping advance the world-changing
research efforts of the Manhattan Project, the foresight of Canadian
scientific leaders such as C. J. Mackenzie, placed Canada among the
global leaders in nuclear research following the war. This legacy of
excellence in nuclear research, continues to this day at ``Canadian
Nuclear Laboratories'' the modern incarnation of the Montreal
Laboratory. In the 79 years since Hans Halban arrived in Montreal,
Canada has developed a large fleet of domestically designed CANDU
heavy-water moderated power reactors which are able to meet a
significant fraction of the nation's power requirements. This
impressive technological achievement is the legacy of a remarkable
collaboration among wartime allies to harness cutting edge scientific
advancements to bring the war to a swift and successful close.

\section*{Acknowledgments}

The authors would like to thank the following colleagues for their
assistance preparing this manuscript: Andrew Prudil, Alan Carr, Daniel
Alcazar, Tom Kunkle and Mark Chadwick. We would also like to thank the
numerous reviewers who provided thorough and helpful feedback on this
paper. This work was supported by the US Department of Energy through
the Los Alamos National Laboratory. Los Alamos National Laboratory is
operated by Triad National Security, LLC, for the National Nuclear
Security Administration of the US Department of Energy under Contract
No. 89233218CNA000001.

\vspace{0.25in}
\noindent\rule{0.35\textwidth}{.4pt}

\bibliographystyle{ans_abbrev}
\small\bibliography{trinity.bib}  

\end{document}